\documentclass[english,twocolumn,showpacs,preprintnumbers,prb,amsmath,amssymb]{revtex4}
\usepackage[T1]{fontenc}
\usepackage[latin1]{inputenc}
\usepackage{textcomp}
\usepackage{amsbsy}
\usepackage{graphicx}

\makeatletter

\makeatletter

\usepackage{dcolumn}

\usepackage{bm}

\makeatother

\makeatother

\usepackage{babel}

\begin{document}

\title{X-ray fluorescence spectroscopy from ions at charged vapor/water interfaces}

\author{Wei Bu and David Vaknin}

\affiliation{Ames Laboratory, and Department of Physics and Astronomy, Iowa State
University, Ames, Iowa 50011, USA}

\date{\today}

\begin{abstract}
X-ray fluorescence spectra from monovalent ions ($\mathrm{Cs^{+}}$)
that accumulate from dilute solutions to form an ion-rich layer near
a charged Langmuir monolayer are presented. For the salt solution
without the monolayer, the fluorescence signals below the critical
angle are significantly lower than the detection sensitivity and only
above the critical angle signals from the bulk are observed. In the
presence of a monolayer that provides surface charges, strong fluorescence
signals below the critical angle are observed. Ion density accumulated
at the interface are determined from the fluorescence. The fluorescent
spectra collected as a function of incident X-ray energy near
the $L_{III}$ edge yield the extended absorption spectra from the
ions, and are compared with recent independent results. The fluorescence
data from divalent $\mathrm{Ba^{2+}}$ with and without monolayer
are also presented.


\end{abstract}

\pacs{61.10.Kw, 61.10.Ht, 73.30.+y 82.45.Mp}

\maketitle

\section{Introduction}

Recently, we reported on the spatial distributions of monovalent ions
($\mathrm{C}\mathrm{s}^{+}$) at highly charged interfaces at $\sim3$
{\AA} resolution by using synchrotron x-ray anomalous reflectivity
techniques \cite{Bu2005,Bu2006}. We demonstrated that these distributions
are well described by a Poisson-Boltzmann theory that accounts for
proton release and binding to a R-PO$_{4}$H group (R is typically
a fatty acid portion of the molecule). Subsequently, we reported on
the extension of these studies by analyzing x-ray energy scans at
fixed momentum-transfers ($Q_{z}$) under specular reflectivity conditions.
In addition to obtaining ion distributions, our analysis yielded the
energy dependence of the dispersion corrections $f'(E)$ and $f''(E)$
near the $\mathrm{C}\mathrm{s}^{+}$ $L_{III}$ resonance \cite{Bu2006a}.
This study confirmed the ion density accumulations at the charged
interfaces and provided spectroscopic information of the ions with
details that shed light on the immediate environment of the ions,
similar to that obtained by extended x-ray absorption fine structure
spectroscopy (EXAFS) experiments.

The x-ray fluorescence near total reflection is another common technique
to determine ion adsorption to charged Langmuir monolayers at air/solution
surface \cite{Bloch1985,Bloch1990,Yun1990,Daillant1991,Novikova2003,Zheludeva2003,Shapovalov2007}.
Herein, we report detailed determination of fluorescence spectra from
monovalent ions $\mathrm{C}\mathrm{s}^{+}$ and divalent ions $\mathrm{Ba^{2+}}$,
both in solutions and as they form an ion-rich layer near the charged
interfaces. We compare the findings with recent results obtained from the
anomalous reflectivity technique \cite{Bu2005,Bu2006}. In the present
study, we extend on previous studies by exploring the fluorescence
signals as a function of photon energies, in particular, near resonances.
As shown below, our approach yields the energy dependence of the dispersion
corrections of $\mathrm{C}\mathrm{s}^{+}$, $f'(E)$ and $f''(E)$,
near a resonance, which is known to be affected by the local non-crystalline
environment of the ions. In the past, such corrections were obtained
by Bijvoet Pairs at Bragg reflections \cite{Templeton1980}, by absorption
cross-section measurements \cite{Kemner1996,Gao2005}, and by calculation
using atomic wave functions \cite{Cromer1970}. In general,
 EXFAS and related spectroscopic experiments are conducted in transmission
configurations, but it is known that fluorescence experiments, as in this study, can yield similar results.

\section{Experimental Setup and Methods}

\begin{figure}[htl]
\includegraphics[width=2.4in]{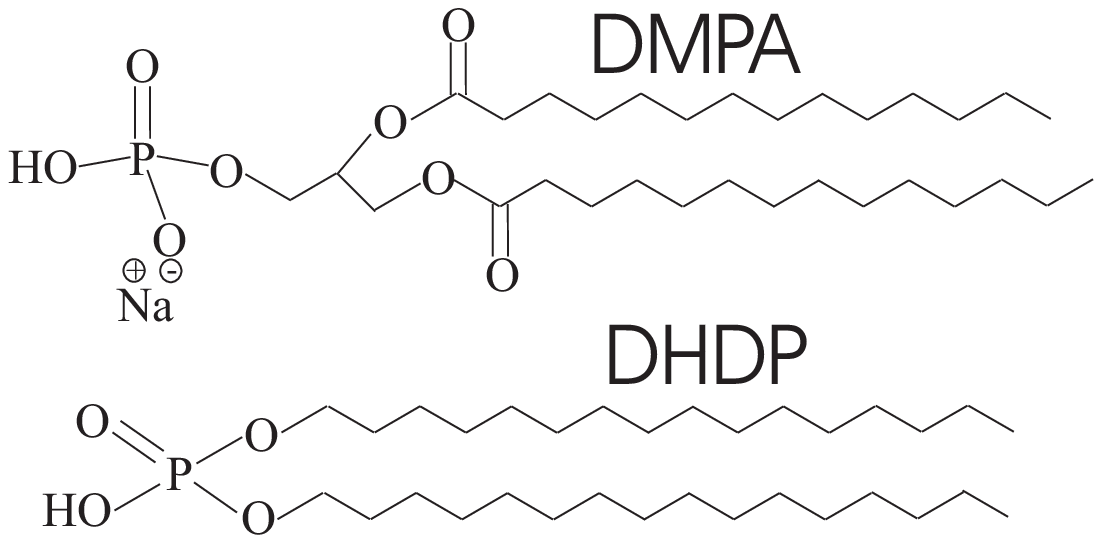}
\caption{\label{molecules} Dimysteroyl phosphatidic acid (DMPA) and dihexadeocyl
hydrogen phosphate (DHDP) molecules used to form the Langmuir monolayers.}
\end{figure}

\begin{figure}[htl]
\includegraphics[width=2.4 in]{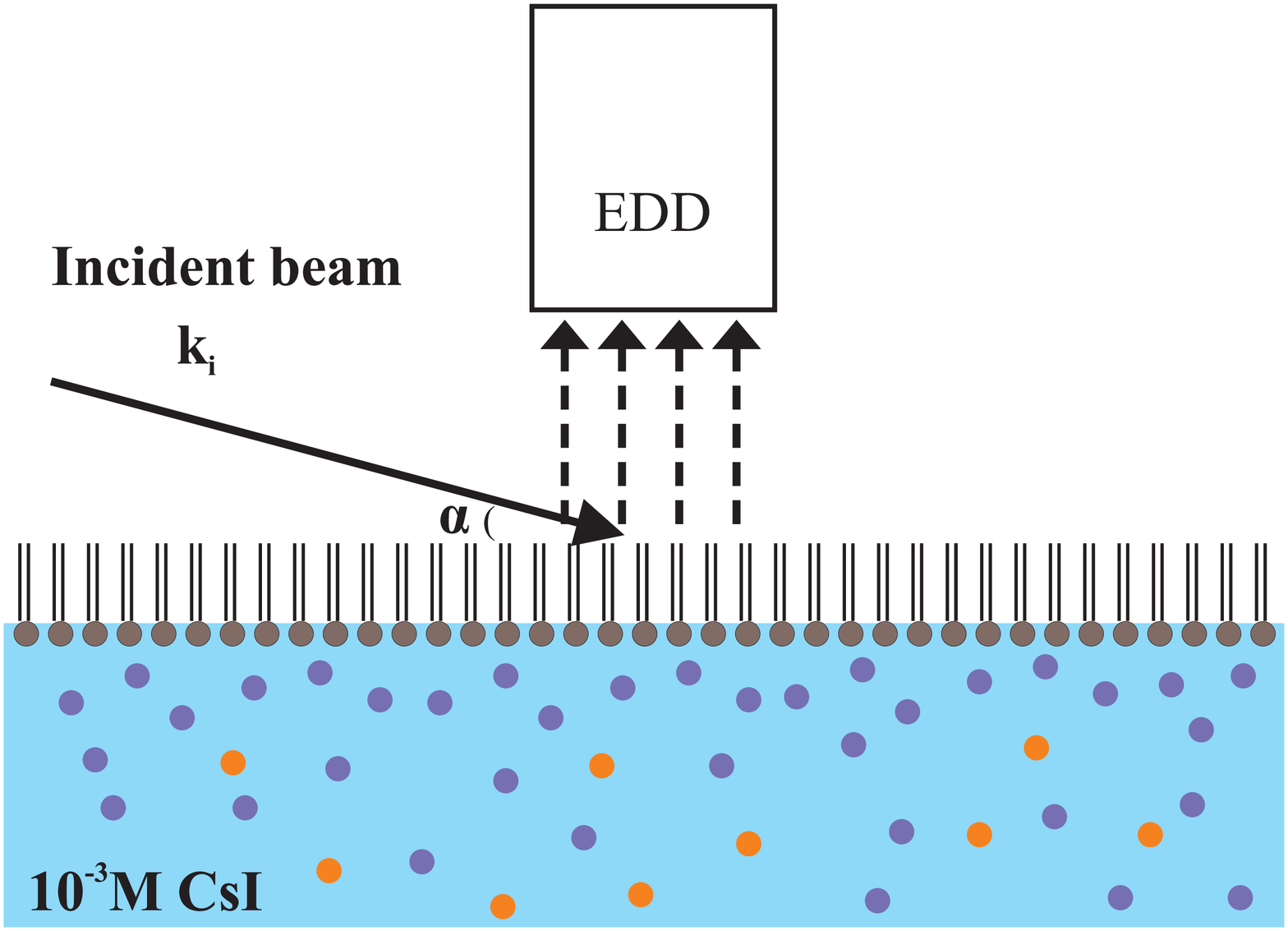}
\caption{\label{setup} Illustration of the fluorescence experiment setup.
The monolayer is spread in an encapsulated Langmuir trough purged with water saturated helium. The Vortex-EX\textregistered{} Multi-Cathode X-Ray Detector window (50 mm$^{2}$ effective detector
area) is placed at a distance $\sim 2$ cm from the surface.  The fluorescent beam goes through a thin Kapton
window that seals the trough. }

\end{figure}

To form well-controlled interfacial charges, monolayers of dimysteroyl
phosphatidic acid (DMPA) and dihexadecyl hydrogen phosphate (DHDP)
(Fig.\ \ref{molecules}) were spread at salt (e.g., $\mathrm{CsI}$
and $\mathrm{BaI_{2}}$) solution/gas interfaces \cite{Vaknin2003,Gregory1997,Gregory1999}.
Detailed procedures of sample preparations and handling were described
elsewhere \cite{Bu2005,Bu2006}. Isotherms have been used for controlling
the molecular area, fixed at $41\pm1$ {\AA}$^{2}$ in
all discussed experiments. \emph{In-situ} fluorescence spectra at
the gas/liquid interface were conducted on the Ames Laboratory Liquid
Surface Diffractometer at the Advanced Photon Source, beam-line 6ID-B
(described elsewhere\cite{Vaknin2003a}). The highly monochromatic
X-ray beam, selected by a downstream Si double crystal monochromator,
is deflected onto the liquid surface at a desired angle of incidence
with respect to the liquid surface by a second monochromator (y-cut
quartz single-crystal \textit{d}-spacing 4.25601 {\AA}) mounted
on the liquid-surface diffractometer yielding energy resolution $\sim0.85$
eV in the vicinity of the $\mathrm{C}\mathrm{s}^{+}$ $L_{III}$ resonances
of the ions ($\sim5$ keV). The absolute scale of the x-ray energy
was calibrated with various absorption edges to better than $\pm2$
eV. The incident photon energy can be continuously varied from 4 to
40 keV, in a fixed-$Q_{z}$ mode, namely, adjusting all angles to
maintain fixed momentum transfer. Scattered photon intensities are
normalized to an incident beam-monitor in front of the sample. Vortex-EX\textregistered{}
Multi-Cathode X-Ray Detector (SII Nano Technology USA, Inc.), an energy-dispersive-detector (EDD),
is lowered to the surface in an aluminum
well with a thin Kapton window located  $\sim2$ cm above the
liquid surface (Fig.\ \ref{setup}). The Langmuir trough is placed
in a sealed canister kept under a flow of water-saturated
helium gas.

Fluorescence is an ion-specific technique in that it can distinguish
contributions from different ions because of their characteristic
fluorescence spectra\cite{Zheludeva2003,Novikova2003,Shapovalov2007,Bu2008}. Since the x-ray penetration depth
changes dramatically (from $60-80$ {\AA} to $1-2\;\mu\mathrm{m}$)
around the critical angle ($Q_{c}\sim0.022$ {\AA}$^{-1}$ for
total reflection), the fluorescence signals below and above the critical
angle for all solutions in the present study are dominated by different
regions of the systems. Below the critical angle, the signal is less sensitive to contributions from the pure bulk solution and in the presence of charges is dominated
by ions at the surface, due to the finite penetration depth of X-rays.
On the other hand, above the critical angle, the fluorescence signals
consist of contributions from the ions in the bulk and at the interface.

\section{Results and Discussion}
\subsection{Surface ion enrichment}

\begin{figure}[h]
\includegraphics[width=2.8in]{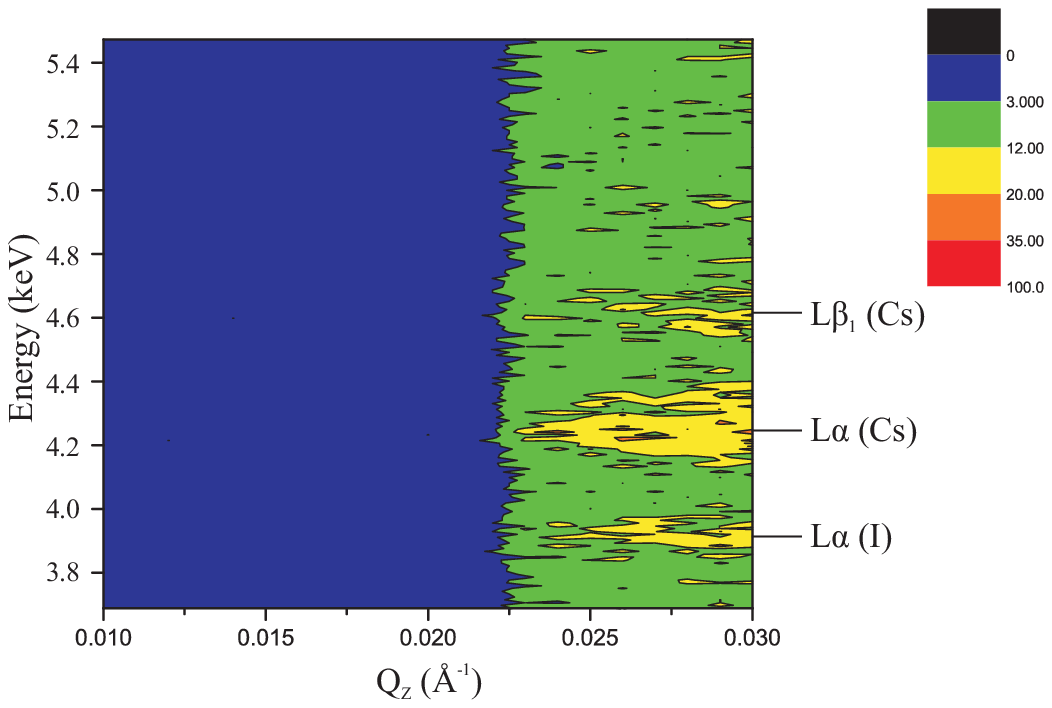}
\includegraphics[width=2.8in]{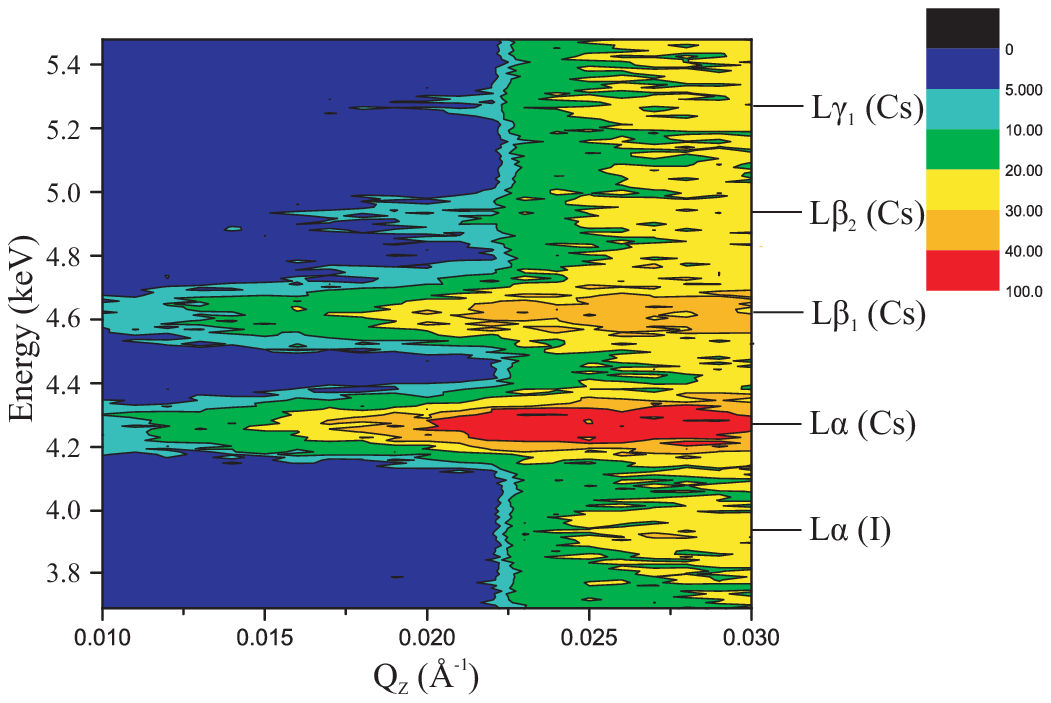}
\caption{\label{8kev_cont} Contour plots of fluorescence intensity for $\mathrm{10^{-3}}$
M CsI without (upper panel) and with monolayer DHDP (lower panel).
Emission lines are labeled on the right side. Incident x-ray beam
energy is 8 keV.}
\end{figure}

\begin{figure}[thl]
\includegraphics[width=2.2 in]{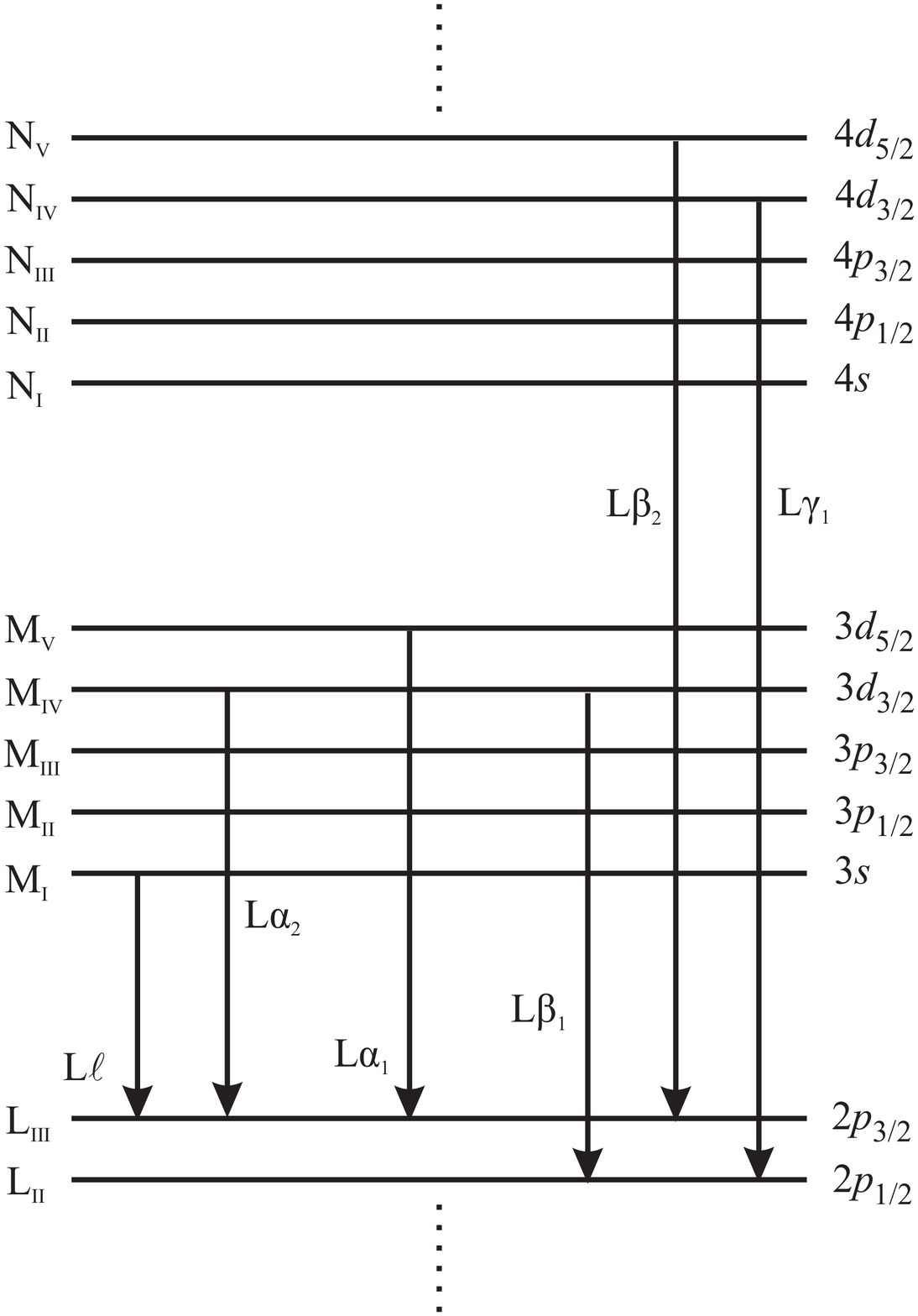}
\caption{\label{spec_diag} Schematic electron energy levels and main transition
processes of the emission lines from the $L$ shell.}
\end{figure}

Figure\ \ref{8kev_cont} shows contour plots of fluorescence intensity
as functions of X-ray photon energy, $E$, and momentum transfer, $Q_{z}$,
for $10^{-3}$ M CsI with and without DHDP monolayer. Without monolayer
(upper panel), the fluorescence pattern is relatively simple. Below
the critical angle ($Q_{z}<Q_{c}$), no significant fluorescence intensity
is observed, consistent with the fact that ions (e.g., $\mathrm{Cs^{+}}$, $\mathrm{I^{-}}$)
in the bulk are not concentrated enough to generate any detectable
intensity over the very short penetration depth. For this ion concentration (10$^{-3}$ M), we can practically claim that ions in the bulk have no contribution
to the fluorescence signal in this $Q_{z}$ range, or that for this
concentration the signal from the surface (below the critical angle)
is significantly lower than the sensitivity of our detector. This
is true at least for dilute concentrations ($10^{-3}$ M), but not for higher concentrations as shown below.
Above the critical angle ($Q_{z}>Q_{c}$), the x-ray beam penetrates much
deeper ($1-2\;\mu\mathrm{m}$), and this concentration is sufficient
to generate fluorescence signals. A few main emission lines from $\mathrm{Cs^{+}}$
($L\alpha$ and $L\beta_{1}$) and $\mathrm{I^{-}}$ ($L\alpha$)
are clearly identified. A general scheme defining the $L$ shell emission lines
is shown in Fig.\ \ref{spec_diag}.
The fluorescence pattern with the DHDP monolayer (lower panel) is significantly
different from that of the bare surface below the critical angle showing
emission lines from $\mathrm{Cs^{+}}$, but none from $\mathrm{I^{-}}$.
These emission lines include a few weaker ones ($L\beta_{2}$ and
$L\gamma_{1}$), not observed from the bulk of the pure
solution (upper panel). This is qualitative evidence that $\mathrm{Cs^{+}}$
exclusively adsorb at the negatively charged surface, and
no emission lines from $\mathrm{I^{-}}$, including the strongest
$L\alpha$, are not observed below the critical angle. 
This is qualitative evidence that $\mathrm{Cs^{+}}$
exclusively adsorb at the negatively charged surface.  No emission lines from
$\mathrm{I^{-}}$; , including the strongest L$\alpha$ , are observed below the critical angle.
This implies that within the uncertainty of our measurement (about 0.1 ions per
DHDP molecule) there is no enrichment of $\mathrm{I^{-}}$ at the interface.
Using DMPA as a monolayer yields essentially the same fluorescence patterns (data not
shown), consistent with theoretical predictions\cite{Travesset2006}.

\begin{figure}[htl]
\includegraphics[width=2.4 in]{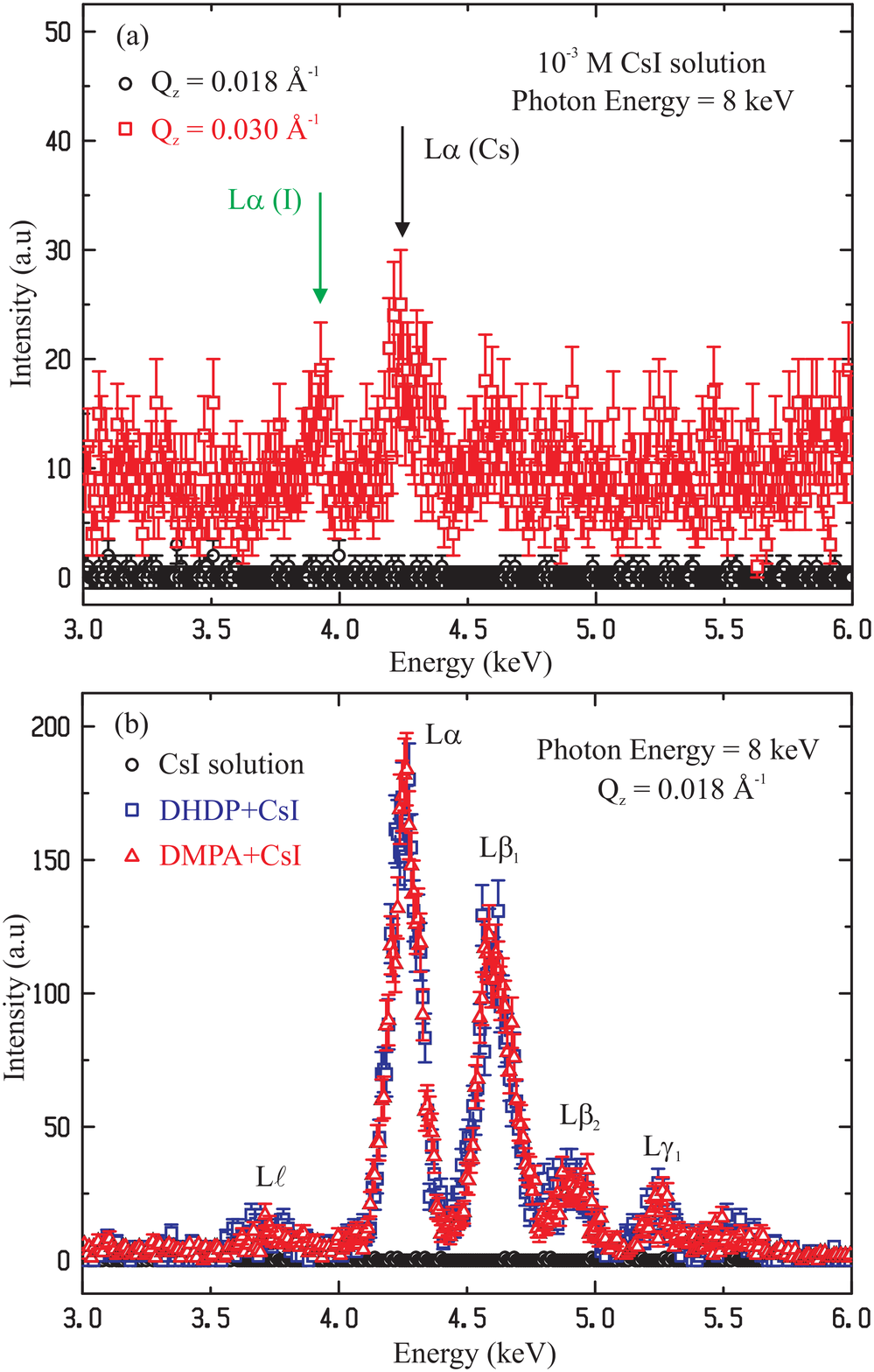}
\caption{\label{8kev_E} (a) Fluorescence intensity versus emission line energy
for $10^{-3}$ M CsI below and above the critical angle as indicated.
(b) Fluorescence intensity versus emission line energy at $Q_{z}=0.018$
{\AA}$^{-1}$ with and without monolayer at the interface. }
\end{figure}

Figure\ \ref{8kev_E} (a) shows $E$-cuts (cuts along the energy
axis at a specific $Q_{z}$ value) of the fluorescence pattern for $10^{-3}$
M CsI without the monolayer below ($Q_{z}=0.018$ {\AA}$^{-1}$)
and above the critical angle ($Q_{z}=0.030$ {\AA}$^{-1}$). Fluorescence
signals are observed only above the critical angel. In the presence of monolayers
(DHDP and DMPA), the $E$-cuts below the critical
angle are shown in Fig.\ \ref{8kev_E} (b). As indicated, the emission
lines from $\mathrm{Cs^{+}}$ are labeled, but no emission lines of
$\mathrm{I}^{-}$ (e.g., $L\alpha$, $\sim3.9$ keV) are detected. The
DHDP or DMPA monolayers have practically identical fluorescence signals,
which implies they have similar amounts of $\mathrm{Cs^{+}}$
ions at the surface. This is theoretically expected, according to
the renormalized Poisson-Boltzmann theory\cite{Bu2006}. This is
because DHDP and DMPA have similar $pK_{\alpha}$ ($\sim2.1$) for
the first proton release. At this concentration, it is not expected
that the second hydrogen in DMPA will be released, unlike in the case
of the divalent\cite{Vaknin2003} or the trivalent ion solutions\cite{Pittler2006}.

\begin{figure}[htl]
\includegraphics[width=2.8in]{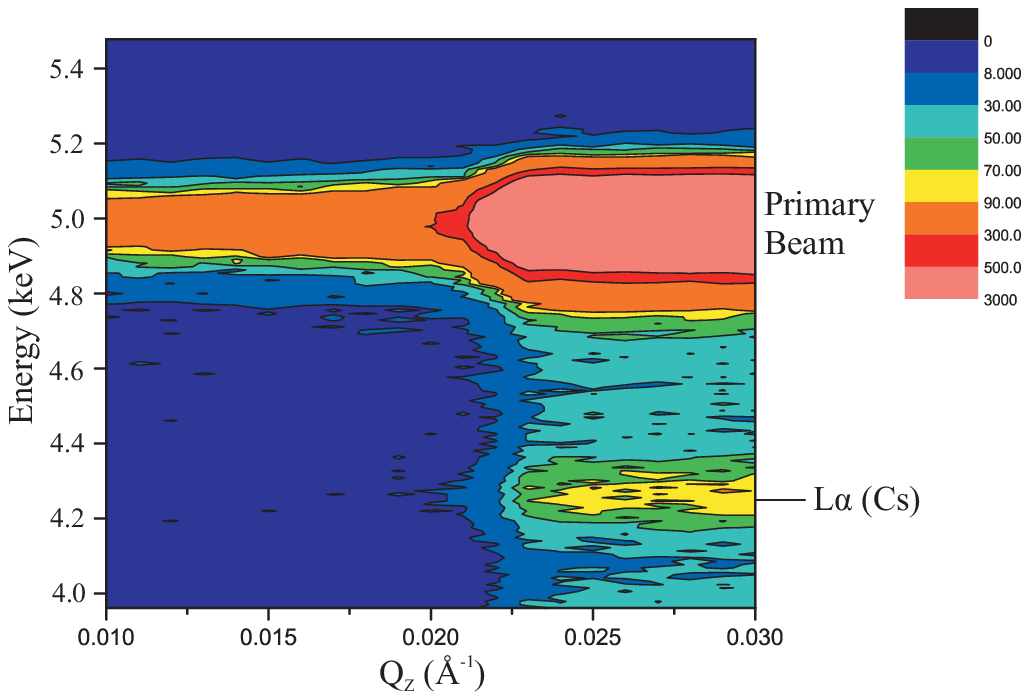}
\includegraphics[width=2.8in]{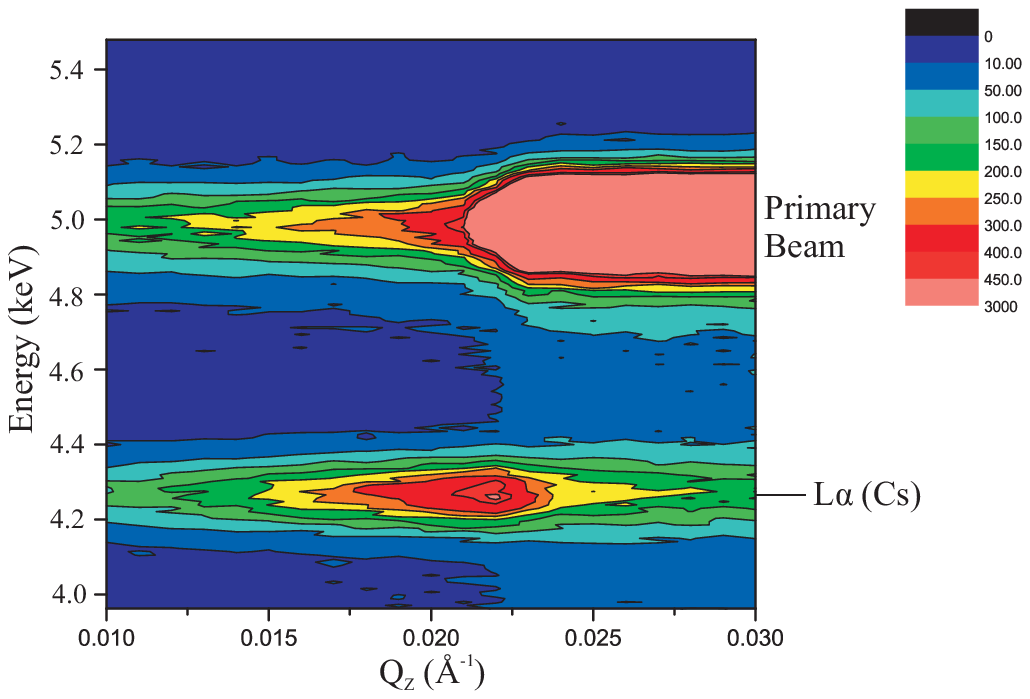}
\caption{\label{5kev_cont} Contour plots of fluorescence intensity for $\mathrm{10^{-3}}$
M CsI without (upper panel) and with monolayer DHDP (lower panel).
Incident x-ray beam energy is 5.015 keV.}
\end{figure}

\begin{figure}[htl]
\includegraphics[width=2.4in]{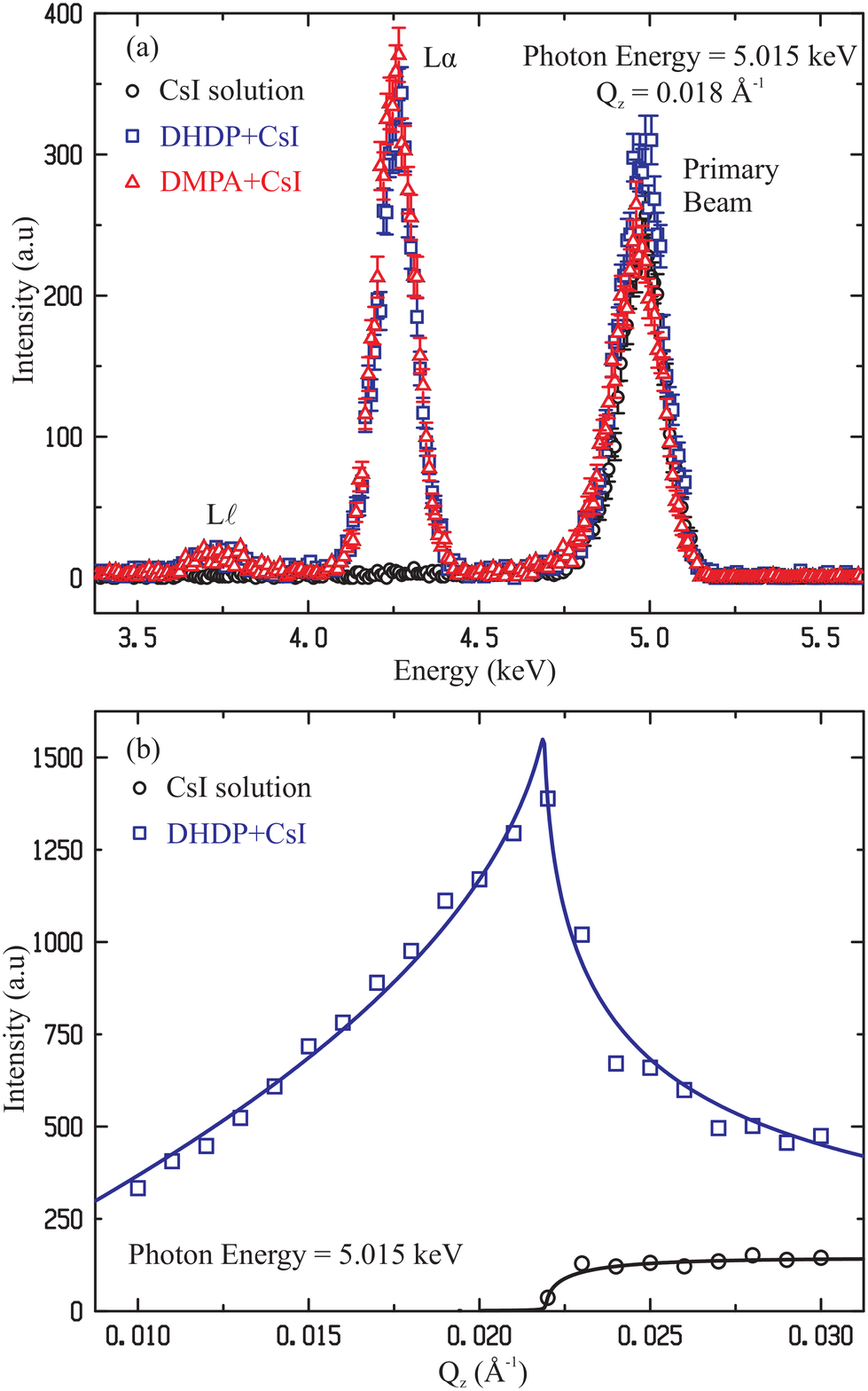}
\caption{\label{5kev_E} (a) Fluorescence intensity versus emission line energy
at $Q_{z}=0.018$ {\AA}$^{-1}$ with and without monolayer materials
($E$-cuts from Fig.\ \ref{5kev_cont}). (b) Fluorescence intensity
of $\mathrm{Cs^{+}}$ $L\alpha$ emission line versus $Q_{z}$ with
and without DHDP ($Q_{z}$-cuts Fig.\ \ref{5kev_cont}).}
\end{figure}

Contour plots of fluorescence intensity for $10^{-3}$ M CsI with
and without the monolayer DHDP with incident x-ray beam energy at
the Cs $L_{III}$ resonance (5.015 keV), are shown in Fig.\ \ref{5kev_cont}.
The strong intensity ridge at approximately 5 keV is due to scattering
of the incident beam, labeled as the primary beam. This signal consists of
primarily elastic and Compton inelastic scattering. Figure\ \ref{5kev_E}(a)
shows $E$-cuts obtained from Fig.\ \ref{5kev_cont} below the critical
angle ($Q_{z}=0.018$ {\AA}$^{-1}$). Because the incident beam
energy is near the $\mathrm{C}\mathrm{s}^{+}$ $L_{III}$ resonance,
only emission lines from $L_{III}$ ($L{\normalcolor l}$, $L\alpha,$) are observed
(the $L\beta_{2}$ is  entangled with the primary beam).
Figure\ \ref{5kev_E}(b) shows the $Q_{z}$-cuts of $\mathrm{Cs^{+}}$
$L\alpha$ emission line from those contour plots. Without a DHDP
or DMPA monolayer, the fluorescence signal is observed only above
the critical angle, that is from the bulk. The intensity slightly increases
with $Q_{z}$ since the penetration depth becomes longer with $Q_{z}$\cite{Yun1990}.
With the DHDP monolayer, fluorescence intensity below the critical
angle, due to surface enrichment of $\mathrm{Cs^{+}}$ at the surface
is observed. This intensity reaches a maximum value at the critical
angle, due to the multiple scattering, as predicted by the distorted
wave Born approximation \cite{Vaknin2003a,Vineyard1982,Kjaer1994}.

\begin{figure}[htl]
\includegraphics[width=2.4 in]{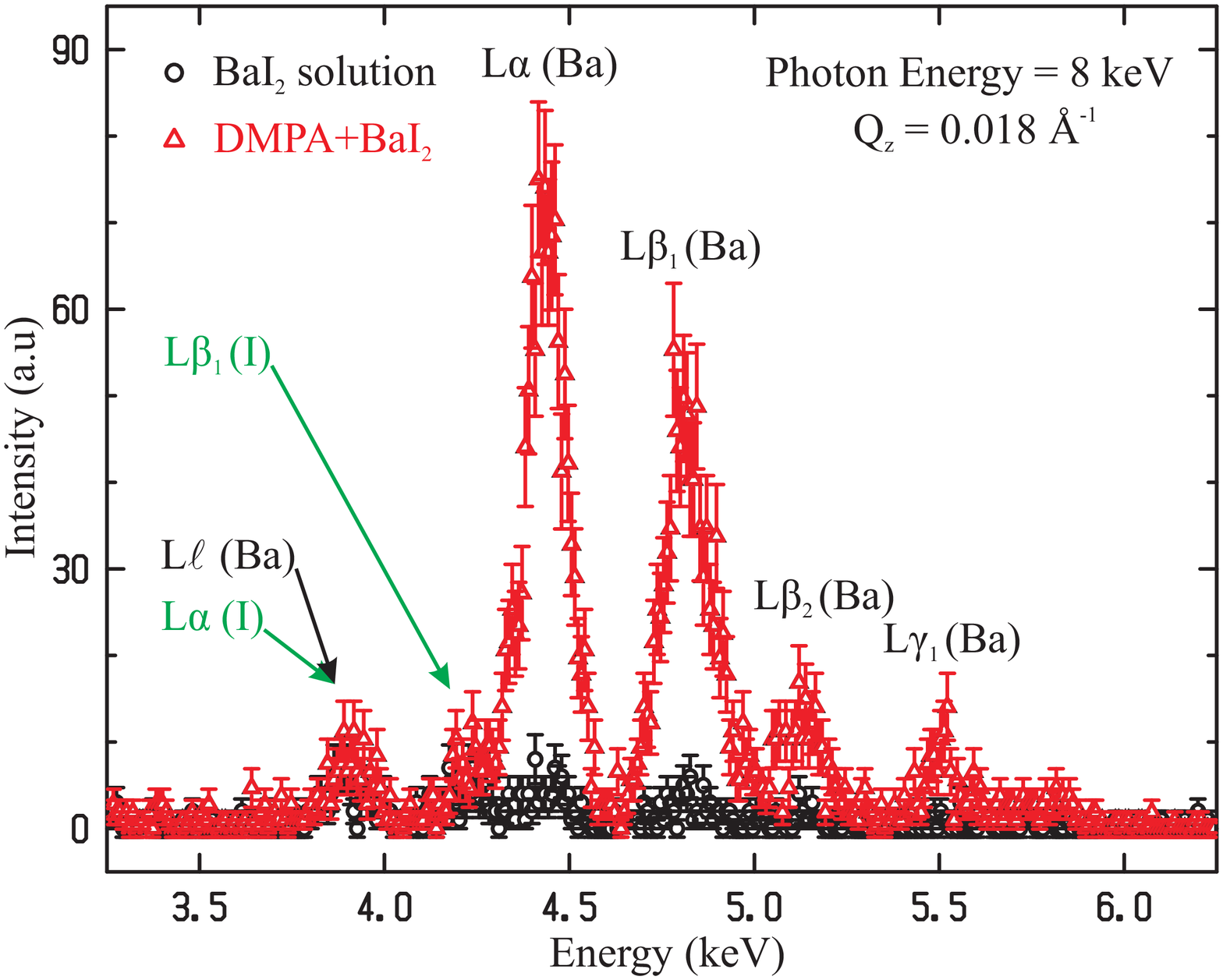}
\caption{\label{BaI} Fluorescence intensity versus emission line energy for
$10^{-2}$ M $\mathrm{BaI_{2}}$ with and without DMPA at $Q_{z}=0.018$
{\AA}$^{-1}$. Emission lines from both $\mathrm{Ba^{2+}}$ and
$\mathrm{I^{-}}$ are labeled.}
\end{figure}

Similar experiments performed with $\mathrm{BaI_{2}}$ solution (for
$10^{-2}$ M) with monolayers produce similar results. The fluorescence
data below the critical angle with and without the DMPA monolayer
are shown in Fig.\ \ref{BaI}. Because of the higher bulk concentration
(than that used with CsI), the emission lines from both $\mathrm{Ba^{2+}}$
and $\mathrm{I^{-}}$ are observed below the critical angle for the
bare surface solution without the monolayer. The presence of DMPA
charges at the interface enhance the Ba emission lines, with no detectable
change in the intensities of the $\mathrm{I^{-}}$ emission lines.

\subsection{Evaluating interfacial ion concentration}

Above the critical angle, the intensity of the X-ray beam, $I(z)$,
decays as it penetrates deeper into the sample as follows, $I(z)=I_{0}e^{-z/D(\alpha)}$,
where $I_{0}$ is the incident beam intensity and $D(\alpha)$ is
the penetration depth, which is a function of the incident beam angle
($\alpha$) and x-ray energy. Assuming the fluorescence intensity
from one ion is $CI_{0}$ ($C$ is a scale factor, determined
from the experimental setup), the fluorescence intensity collected
by the detector has two contribution from the surface and bulk, $I_{s}$
and $I_{b}$, respectively. The surface scattering is given by, \begin{equation}
I_{s}=CI_{0}AN_{ion}/A_{lipid},\label{I_sur}\end{equation}
 and the bulk scattering is \begin{equation}
I_{b}=CI_{0}A\rho_{bulk}\int_{0}^{\infty}e^{-z/D(\alpha)}dz=CI_{0}A\rho_{bulk}D(\alpha),\label{I_bulk}\end{equation}
 where $A$ is the detector area, $N_{ion}$ is the number of ions
per lipid, $A_{lipid}$ is the lipid area (i.e., molecular area),
and $\rho_{bulk}$ is the ion bulk concentration. $I_{b}$ can be
obtained from the fluorescence data of the pure solution without the
monolayer, while $I_{s}$ can be obtained from the fluorescence data
of the solution with the monolayer after the subtraction of $I_{b}$.
Using Eqs. \ (\ref{I_sur}) and \ (\ref{I_bulk}), one can readily
get the number of ions per lipid at the surface, using the following
relation, \begin{equation}
N_{ion}=\frac{I_{s}(\alpha)}{I_{b}(\alpha)}A_{lipid}D(\alpha)\rho_{bulk}.\label{num_ions}\end{equation}
The absorption of emitted photons as they traverse to the EDD is negligible,
since their path in the sample is shorter than that of the incident
beam by a factor of at least 100.

\begin{figure}[htl]
\includegraphics[width=2.4in]{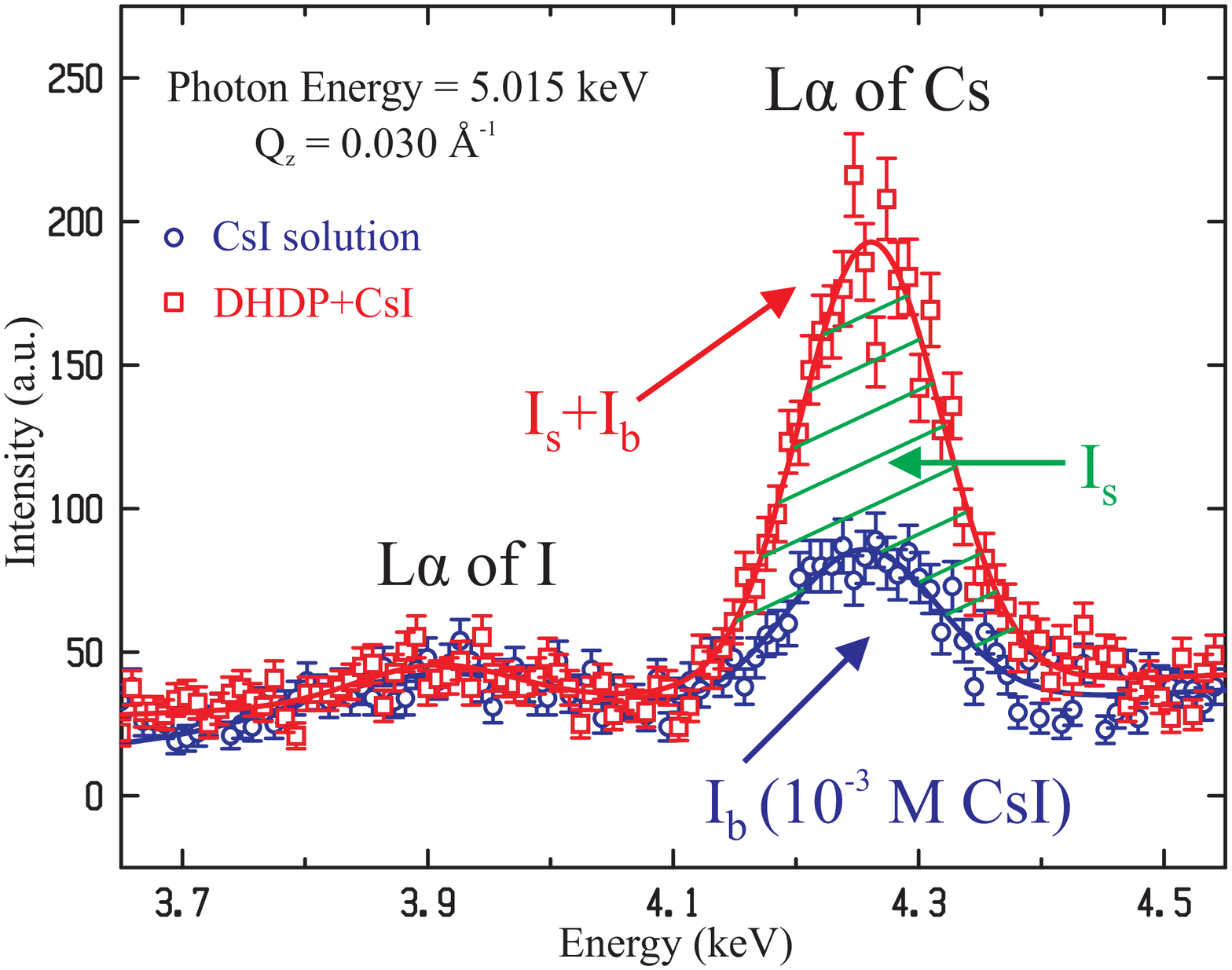}
\caption{\label{Escan_quant} Fluorescence data above the critical angle ($Q_{z}=0.030$
{\AA}$^{-1}$) for $10^{-3}$ M CsI with ($I_{s}+I_{b}$) and without
DHDP ($I_{b}$). The shaded area represents $I_{s}$ of $\mathrm{Cs^{+}}$
$L\alpha$ emission line. X-ray energy is 5.015 keV.}
\end{figure}

Figure\ \ref{Escan_quant} shows that the fluorescence data, above
the critical angle, at $Q_{z}=0.030$ {\AA}$^{-1}$ for $10^{-3}$
M CsI with and without the monolayer DHDP, and the definition of measured
$I_{s}$ and $I_{b}$. Spreading a DHDP monolayer enriches the surface
with $\mathrm{Cs^{+}}$ ions and enhances the $L\alpha$ emission
line (shaded area), but does not change the intensity of $\mathrm{I^{-}}$
$L\alpha$ emission line. Some of the emission lines from different ions overlap, due to the poor resolution of the EDD (150 - 200 eV). For instance, the emission line of the $\mathrm{Cs^{+}}$
$L\alpha$ line (4.3 keV), and the $\mathrm{I^{-}}$ $L\beta_{1}$  may cause overestimating $I_{b}$ from the Cs line.  To overcome this problem, we used the PyMca program
to fit fluorescent data to obtain the relative contributions of both lines from Cs and I
\cite{Pymca}. As shown in Eq. \ (\ref{num_ions}), the number of
ions can be obtained by evaluating $I_{s}$ and $I_{b}$ at any $Q_{z}$
above the critical angle.  In this study, $I_{s}$ and $I_{b}$
were measured at eight different $Q_{z}$ values (from 0.023 to 0.030
{\AA}$^{-1}$), yielding an average $0.47\pm0.09$ and $0.54\pm0.09$
$\mathrm{Cs^{+}}$ per lipid for $10^{-3}$ M CsI solution with DHDP
and DMPA as a monolayer, respectively. Both values are in an good
agreement with anomalous reflectivity\cite{Bu2006} and constant-$Q_z$ energy scans\cite{Bu2006a} studies,
where more complicated data analysis is required.

\subsection{Evaluating the Fine Structure $\mathbf{\boldsymbol{f^{\prime\prime}(E)}}$}
The intensity of the emission line is proportional to the absorption
of the ion, which is strongly dependent on photon energy near an absorption
edge, and also the immediate environment of the ion. By varying
the incident beam energy at a fixed $Q_{z}$, we obtain the energy
dependence of the absorption correction, namely $f^{\prime\prime}(E)$
up to a scale factor. Performing this experiment below the critical
angle does not require any geometry or absorption corrections,
since there is negligible bulk contribution to the signal as the emitting ions are concentrated at the first 10 {\AA}
of the surface\cite{Bu2006}.

\begin{figure}[htl]
\includegraphics[width=2.6 in]{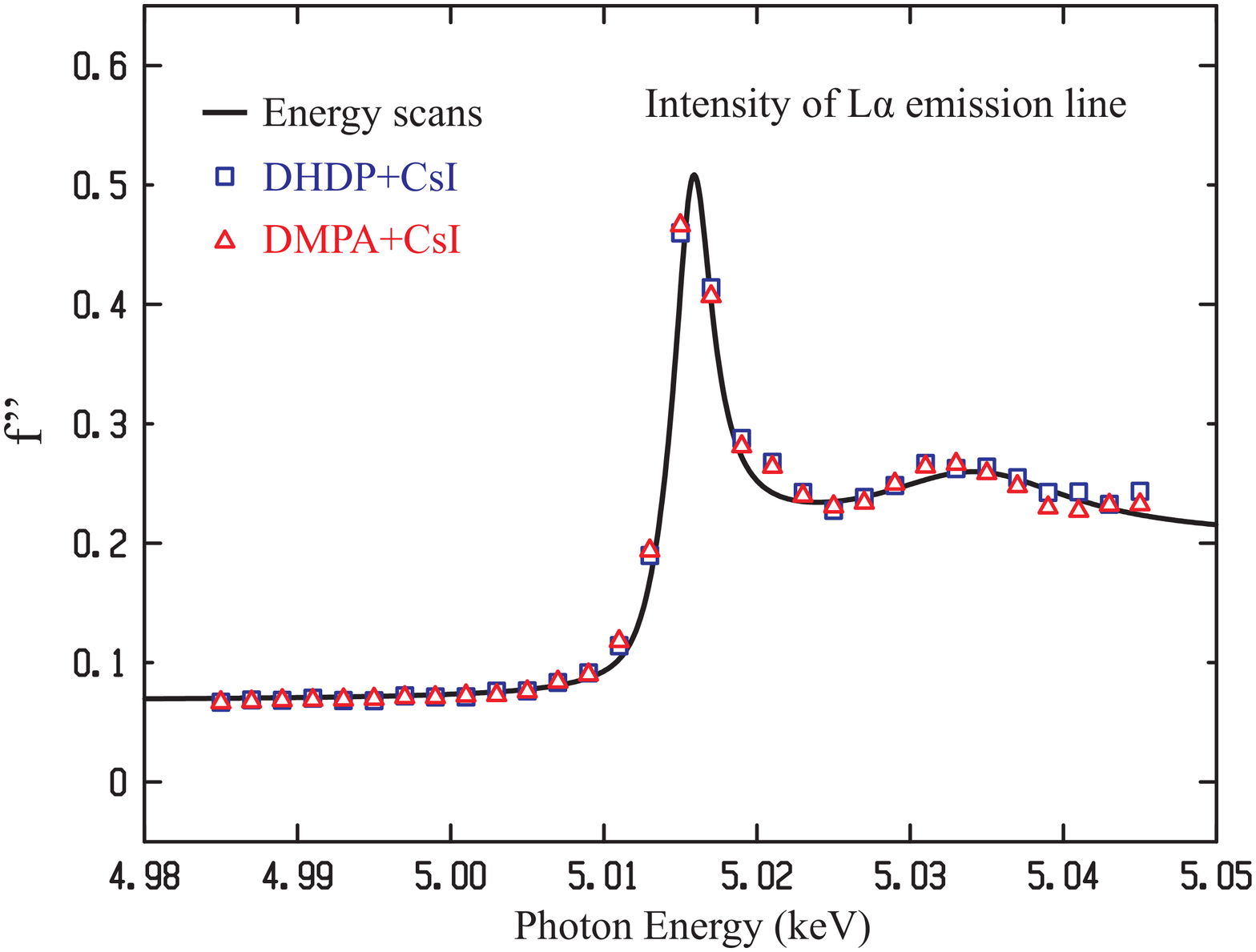}
\caption{\label{EXAFS} Fluorescence intensity of the $\mathrm{Cs^{+}}$ $L\alpha$
emission line from $10^{-3}$ M CsI with the monolayers (DHDP and
DMPA) at $Q_{z}=0.018$ {\AA}$^{-1}$. The intensity is scaled
to the $f^{\prime\prime}(E)$ far away from the $\mathrm{C}\mathrm{s}^{+}$
$L_{III}$ resonance ($\pm30$ eV). The solid line is $f^{\prime\prime}(E)$
obtained by fixed-$Q_z$ energy scans \cite{Bu2006a}.}
\end{figure}

Figure\ \ref{EXAFS} shows the fluorescence intensity of the $\mathrm{Cs^{+}}$
$L\alpha$ emission line as a function of the incident beam energy.
DHDP and DMPA were used as monolayer materials, $Q_{z}$ was fixed
at $0.018$ {\AA}$^{-1}$ to minimize the bulk contribution. The
incident beam energy was scanned around the $\mathrm{C}\mathrm{s}^{+}$
$L_{III}$ resonance. Far away from that resonance ($\pm30$ eV),
$f^{\prime\prime}(E)$ for Cs is known from various
experimental and theoretical studies\cite{Cromer1970}.
In the vicinity of the resonance, $f^{\prime\prime}(E)$
of the emitting ion can be influenced by the local
environment, and the spectra becomes more complex.
By scaling the $f^{\prime\prime}(E)$
values away from the resonance ($\pm30$ eV), the fluorescence intensity
of the $\mathrm{Cs^{+}}$ $L\alpha$ emission line can be converted
to the specific $f^{\prime\prime}(E)$ in its interfacial environment.
As shown in Fig.\ \ref{EXAFS}, the fluorescence intensity after
scaling agrees well with $f^{\prime\prime}(E)$, obtained in
our previous study\cite{Bu2006a} (solid line) by a more complicated
analysis of constant-$Q_{z}$ energy scans. We note that the reported measurement of $f^{\prime\prime}(E)$ in bulk aqueous environment (see Fig. 5 in Ref.\ \cite{Gao2005}) is slightly different than the one we report here for the $\mathrm{C}\mathrm{s}^{+}$ $L_{III}$ edge.  These differences may arise from the fact that the ions in our study reside at the interface with a slightly different environment than that in bulk solution.

\section{Summary}
In the present study we extended previous x-ray fluorescence
techniques that had been used to determine the interfacial ion enrichment at
charged monolayers\cite{Bloch1985,Bloch1990,Yun1990,Daillant1991,Novikova2003,Zheludeva2003,Shapovalov2007} by tracing all
emissions lines that can be resolved by our EDD.  Although x-ray fluorescence has been
used for detecting the number density of ions of similar systems
by similar means, these either needed complicated data analysis
\cite{Yun1990} or were limited to providing the number density of ions relative to a known
density of another ion at the interface\cite{Shapovalov2007,Bu2008}.
We confirmed that the fluorescence technique below the
critical angle provides a quick and reliable determination
of the presence of ions, specifically by identifying the
characteristic emission lines of each element that fluoresce. We
demonstrated how to calculate the number density of $\mathrm{Cs^{+}}$ ions at the surface
by measuring the fluorescence signals with and without the monolayer above the critical angle.
We have also shown that the fine structure of the absorption
$f^{\prime\prime}(E)$ for the specific ions at the surface can be
readily obtained from fluorescence signals measured as a function
of photon energy near an absorption edge.

The spatial resolution of ion distributions near charged objects
(for instance, membranes, DNA filaments, vesicles, polyelectrolytes, and others)
in aqueous environment have been improved in recent years with the advances in synchrotron
x-ray radiation and have been expanded to more complex systems\cite{Luo2006,Giewekemeyer2007}.
These experimental tools, together with theoretical advances\cite{Andelman2006,Koelsch2007}, brought
new insight into the nature of ion accumulation near charged interfaces that allow distinguishing
between purely electrostatic attraction and ion-specific binding.
The fluorescence technique, first introduced by Bloch and coworkers\cite{Yun1990}, is yet another
independent technique that can identify enrichment of ions, specifically from mixed salt solutions,
and shed light on the local environment of the ions at the interface in a similar manner
to the newly introduced fixed-$Q_z$ energy scan near ion resonances method\cite{Bu2006a,Park2005}.

\begin{acknowledgments}
We thank D. S. Robinson for technical support at the 6-ID beamline.
Ames Laboratory and the MUCAT sector at the APS are supported by the
U.S. DOE, Basic Energy Sciences, Office of Science, under contract
under Contract No. DE-AC02-07CH11358. Use of the Advanced Photon Source
was supported by the U. S. Department of Energy, Office of Science,
Office of Basic Energy Sciences, under Contract No. DE-AC02-06CH11357.
\end{acknowledgments}

\end{document}